\documentclass[12pt]{article}
\usepackage{amssymb}
\usepackage{myart}

\normalbaselines
\input tcilatex
\begin{document}

\title{Deformation principle and further geometrization of physics }
\author{Yuri A.Rylov}
\date{Institute for Problems in Mechanics, Russian Academy of Sciences,\\
101-1, Vernadskii Ave., Moscow, 119526, Russia.\\
e-mail: rylov@ipmnet.ru\\
Web site: {$http://rsfq1.physics.sunysb.edu/\symbol{126}rylov/yrylov.htm$}\\
or mirror Web site: {$http://gasdyn-ipm.ipmnet.ru/\symbol{126}%
rylov/yrylov.htm$}}
\maketitle

\begin{abstract}
The space-time geometry is considered to be a physical geometry, i.e. a
geometry described completely by the world function. All geometrical
concepts and geometric objects are taken from the proper Euclidean geometry.
They are expressed via the Euclidean world function $\sigma _{\mathrm{E}}$
and declared to be concepts and objects of any physical geometry, provided
the Euclidean world function $\sigma _{\mathrm{E}}$ is replaced by the world
function $\sigma $ of the physical geometry in question. The set of physical
geometries is more powerful, than the set of Riemannian geometries, and one
needs to choose a true space-time geometry. In general, the physical
geometry is multivariant (there are many vectors $\mathbf{Q}_{0}\mathbf{Q}%
_{1}$, $\mathbf{Q}_{0}\mathbf{Q}_{1}^{\prime }$,... which are equivalent to
vector $\mathbf{P}_{0}\mathbf{P}_{1}$, but are not equivalent between
themselves). The multivariance admits one to describe quantum effects as
geometric effects and to consider existence of elementary particles as a
geometrical problem, when the possibility of the physical existence of an
elementary geometric object in the form of a physical body is determined by
the space-time geometry. Multivariance admits one to describe discrete and
continuous geometries, using the same technique. A use of physical geometry
admits one to realize the geometrical approach to the quantum theory and to
the theory of elementary particles.
\end{abstract}

\section{Introduction}

Geometrization is the principal direction of the contemporary theoretical
physics development. It began in the nineteenth century. One can list the
following stages of the physics geometrization:

1. Conservation laws of energy-momentum and angular momentum

2. The first modification of the space-time geometry (geometrization of the
space-time, the concept of simultaneity, geometrization of particle motion,
problem of high velocities)

3. The second modification of the space-time geometry (existence of
nonhomogeneous space-time geometry, influence of the matter distribution on
the space-time geometry)

4. Geometrization of charge and electromagnetic field. (Kaluza, O.\ Klein)

5. The third modification of the space-time geometry (the new space-time
geometry of microcosm, the concept of multivariance, geometrization of mass,
existence of geometrical objects in the form of physical bodies, geometrical
approach to the elementary particles theory).

Now the theoretical physics stands before the third modification of the
space-time geometry, connected with investigation of microcosm. The
conventional space-time geometry is insensitive to structure of the
microcosm. From viewpoint of the conventional geometry it is of no
importance, whether the microcosm geometry is discrete or continuous. It is
of no importance, whether geometrical objects may be divided into parts at
no allowance, or their divisibility is restricted. The mathematical
technique of contemporary theoretical physics is based on a use of the
infinitesimal calculus, which supposes continuity of space-time and
unlimited divisibility of geometrical objects. Furthermore, there exists no
effective method, which admits one to construct a discrete geometry, or a
geometry with a limited divisibility. This is connected with the fact that
the contemporary geometry ignores the concept of multivariance and ejects
the concept of multivariance, if it meets accidentally. The third
modification of the space-time geometry is connected with appearance of a
new method of the geometry construction, which describes multivariant
geometries. The multivariant geometry may describe both discrete and
continuous geometries, as well as geometries with a limited divisibility of
geometrical objects \cite{R2006}. These properties appear to be important in
the space-time geometry of microcosm.

Necessity of the third modification appeared in the thirtieth of the
twentieth century, when diffraction of electrons on the small hole was
discovered. Motion of a free particle depends only on the space-time
geometry, and one needs such a space-time geometry, where the free particle
motion be multivariant, and the multivariance intensity depend on the
particle mass. Neither physicists, nor mathematicians could imagine such a
space-time geometry. As a result the problem of the particle motion
multivariance has been solved in the framework of dynamics (but not on the
level of geometry). Classical principles of dynamics in microcosm were
replaced by quantum ones. The problem of multivariant motion of
microparticles has been solved on the level of dynamics. W.Heisenberg
suggested to replace conventional dynamic variables by matrices.
Introduction of matrices is an introduction of a multivariance. However, it
is a multivariance on the level of dynamics. The space-time geometry
remained to be former.

Impossibility of the multivariance problem solution on the geometric level
was connected with imperfection of the method of the geometry construction.
It does not admit one to construct multivariant geometries, which possess
properties, necessary for explanation quantum effects and other properties
of microcosm. Besides, in that time the researchers were under the
impression of advances of quantum mechanics, and multivariance of the
electron motion had been explained as a quantum effect.

In the end of the twentieth century a more perfect method of the space-time
geometry construction has been suggested \cite{R90}. This method is known as
the deformation principle. Geometries, constructed by this method, are known
as tubular geometries (T-geometries) This method is simpler and more
general, than the conventional Euclidean method, because it does not use
such a constraint of the conventional method, as absence of multivariance.
In particular, in the framework of the Riemannian geometry there is only one
plane uniform isotropic space-time geometry: the Minkowski geometry, whereas
in the framework of T-geometries there is a set of plane uniform isotropic
space-time geometries, labelled by a function of one argument. All
geometries of this set (except for the Minkowski one) are multivariant with
respect to timelike vectors. Multivariance of the space-time geometry with
respect to timelike vectors means that there exist vectors $\mathbf{Q}_{0}%
\mathbf{Q}_{1}$, $\mathbf{Q}_{0}\mathbf{Q}_{1}^{\prime }$,... which are
equivalent to timelike vector $\mathbf{P}_{0}\mathbf{P}_{1}$, but not
equivalent between themselves.

As far as there exist many uniform isotropic space-time geometries, we are
to choose the true space-time geometry from this set. We may not choose the
Minkowski space-time geometry on the ground, that this geometry was used
before. We are to make the best of agreement of the space-time geometry with
the experimental data. It appears that the parameters of the space-time
geometry can be chosen in such a way, that the classical principles of
dynamics describe correctly both quantum and classical motion of a free
particle. Of course, the parameters of the true space-time geometry contain
the quantum constant $\hbar $. In this case we do not need the quantum
principles, which in the conventional theory compensate influence of the
incorrectly chosen space-time geometry of microcosm.

Such an expansion of the space-time geometry capacities is connected with
the non-Euclidean method of the geometry construction. This method of the
geometry construction may be qualified as the deformation principle, because
any physical geometry can be obtained as a result of a deformation of the
proper Euclidean geometry. Capacities of the space-time geometries
constructed by means of the deformation principle do not exhausted by
explanation of quantum effects. Structure of elementary particles, their
masses, appearance of short-range force fields in microcosm and such an
enigmatic phenomenon as confinement can be easily explained in terms of the
space-time geometry and its particularity. At any rate the mathematical
technique of T-geometry admits this. In this paper we shall not try to
determine the concrete form of the microcosm geometry. To choose the
concrete microcosm geometry one needs very careful analysis of experimental
data. It is a very difficult problem. We shall show only that mathematical
capacity of microcosm geometry are larger, than that of the contemporary
theory of elementary particles.

The main advantage of the microcosm geometry is the circumstance that it
does not use any hypotheses. Of course, to obtain a concrete space-time
geometry of microcosm, we are to use experimental data and make some
suppositions on the space-time geometry. However, these suppositions will be
made in framework of fixed principles. One may choose only the form of the
world function, which determines the space-time geometry. The principles of
the geometry construction remain to be changeless. They are not a result of
a fitting. They are obtained by means of logic reasonings. This is an
essential difference from the contemporary methods of the elementary
particle theory, where the unprincipled fitting dominates.

Note, that probabilistic and noncommutative geometries are not geometries in
the exact sense of this word. These "geometries" are fortified geometries,
i.e. geometries, equipped with some additional structures (probabilistic and
matrix), given on the Minkowski manifold. In other words in the framework of
these "geometries" the physical geometry, as a science on mutual disposition
of geometric objects remains to be the former geometry of Minkowski. One
adds to this geometry additional structures, introducing multivariance,
which is necessary for the microcosm description. However, this
multivariance is introduced on the dynamic level, but not on the level of
geometry/

Thus, in this paper we demonstrate only mathematical capacity of space-time
geometry in explanation of the microcosm phenomena.

\section{Approaches to geometry}

There are two approaches to geometry. According to the conventional approach
a geometry (axiomatic one) is constructed on a basis of some axiomatics. All
propositions of the axiomatic geometry are obtained from several primordial
propositions (axioms) by means of logical reasonings. Examples of axiomatic
geometries: Euclidean geometry, affine geometry, projective geometry etc.
The main defect of the axiomatic geometry: impossibility of axiomatization
of nonhomogeneous geometries.

Axiomatization of geometry means that from the set $\mathcal{S}$ of all
geometry propositions one can separate several primordial propositions $%
\mathcal{A}$ (axioms) in such a way that all propositions $\mathcal{S}$ can
be obtained from axioms $\mathcal{A}$ by means of logical reasonings.
Possibility of axiomatization is a hypothesis. Its validity has been proved
only for the proper Euclidean geometry \cite{H30}. For nonhomogeneous
geometries (for instance, for the Riemannian one) a possibility of
axiomatization was not proved. In general, a possibility of axiomatization
for nonhomogeneous geometries seems to be doubtful. Felix Klein \cite{K37}
assumed that the Riemannian geometry (nonhomogeneous one) is rather a
geography or a topography, than a geometry.

According to another approach a geometry (physical geometry) is a science on
mutual position of geometrical objects in the space or in the space-time.
(Euclidean geometry, metric\ geometry). All relations of the metric geometry
are finite, but not differential, and the metric geometry may be given on an
arbitrary set of points, but not necessarily on a manifold. It is supposed
that the mutual position of geometrical objects is determined, if the
distance (metric) between any two points of the set is given. The simplicity
of the geometry characteristic and absence of constraints on the set of
points (on the space) is the principal advantage of the metric geometry. The
main defect of metric geometry is its poverty. Such important concepts of
Euclidean geometry as scalar product of vectors and concept of linear
dependence are absent in the metric geometry. However, the proper Euclidean
geometry is a special case of the metric geometry. It means that in the case
of Euclidean geometry the scalar product, concept of linear dependence of
vectors and other concepts and objects of Euclidean geometry can be
expressed via Euclidean metric. These expressions of the Euclidean concepts
via metric are declared to be valid for any metric geometry. Replacing
Euclidean metric by the metric of the metric geometry in question, we obtain
a system of geometrical concepts in any metric geometry. As a result we
obtain the metric geometry, equipped by all concepts of Euclidean geometry 
\cite{R01}.

Besides, one removes such constraints on the metric, as the triangle axiom
and positivity of metric $\rho $. They are not necessary, if concepts of the
Euclidean geometry are introduced in the metric geometry. Instead of metric $%
\rho $ we use the world function $\sigma =\frac{1}{2}\rho ^{2}$, which is
real even in the geometries with indefinite metric (for instance, in the
geometry of Minkowski). We shall refer to such geometries as the tubular
geometries (T-geometry). Such a name of geometry is connected with the fact,
that the straight in T-geometry is a tube, but not a one-dimensional line.
The tubular character of straights in T-geometry is conditioned by the
property of multivariance. Multivariance of a T-geometry means, that there
exist many vectors $\mathbf{Q}_{0}\mathbf{Q}_{1},$ $\mathbf{Q}_{0}\mathbf{Q}%
_{1}^{\prime }$, $\mathbf{Q}_{0}\mathbf{Q}_{1}^{\prime \prime }$, ...which
are equivalent (equal) to vector $\mathbf{P}_{0}\mathbf{P}_{1}$, but are not
equivalent between themselves.

At first sight the multivariance is unexpected and undesirable property of
geometry. However, multivariance is very important in application of
geometry to physics. For instance, the real space-time geometry appears to
be multivariant. In particular, multivariance of the space-time geometry
explains freely quantum effects as geometrical effects. Besides, the
multivariance admits one to set the problem of existence of geometrical
objects in the form of physical bodies. This problem cannot be set in
framework of the Riemannian geometry. (More exact, this problem can be set,
but its solution leads to the statement, that any geometric object can be
realized in the form of a physical body). The statement of this problem is
important, to obtain a simple geometrical approach to the elementary
particles theory.

In T-geometry all geometrical propositions have ready-made form. In
T-geometry there are no theorems, all theorems have been proved in the
Euclidean geometry. Statements of the Euclidean geometry turn into
definitions of T-geometry. All this is rather unusual and is difficult for
perception, because some axioms of Euclidean geometry are not valid in
T-geometry (for instance, the Euclidean axiom: "the straight has no
thickness" is not valid in T-geometry, in general).

To overcome defects of physical and axiomatic geometries we use the fact,
that the proper Euclidean geometry is the axiomatic and physical geometry
simultaneously. The proper Euclidean geometry has been constructed as an
axiomatic geometry, and consistency of its axioms has been proved. On the
other side, the proper Euclidean geometry is a physical geometry and, hence,
it is to be described completely in terms of the metric $\rho $. Indeed,
such a theorem has been proved \cite{R01}.

It is more convenient to introduce the world function $\sigma =\frac{1}{2}%
\rho ^{2}$ instead of the metric $\rho $, because in this case one may
describe the Minkowski geometry and other geometries with indefinite metric
tensor in terms of real world function $\sigma $.

As soon as the proper Euclidean geometry $\mathcal{G}_{\mathrm{E}}$ is a
known geometry, all propositions $P_{\mathrm{E}}$ of the Euclidean geometry $%
\mathcal{G}_{\mathrm{E}}$ can be presented in terms of the world function $%
\sigma _{\mathrm{E}}$ of the proper Euclidean geometry: $P_{\mathrm{E}}=P_{%
\mathrm{E}}\left( \sigma _{\mathrm{E}}\right) $. Replacing the world
function $\sigma _{\mathrm{E}}$ of $\mathcal{G}_{\mathrm{E}}$ by the world
function $\sigma $ of another physical geometry $\mathcal{G}$ in all
propositions $P_{\mathrm{E}}\left( \sigma _{\mathrm{E}}\right) $ of the
Euclidean geometry: $P_{\mathrm{E}}\left( \sigma _{\mathrm{E}}\right)
\rightarrow P_{\mathrm{E}}\left( \sigma \right) $, one obtains all
propositions of the physical geometry $\mathcal{G}$. Replacement of the
world function $\sigma _{\mathrm{E}}$ by other world function $\sigma $
means a deformation of the Euclidean geometry (Euclidean space). It may be
interpreted in the sense, that any physical geometry is a result of a
deformation of the proper Euclidean geometry.

It is very important that all expressions of concepts of the Euclidean
geometry via the world function have a finite (but not differential) form.
The differential form of relations needs an additional information (initial
or boundary conditions). The finite expressions do not need such an
additional information. Besides, using the finite form of relations, we need
to solve some algebraic equations, whereas, using a differential form of
relations we are forced to solve differential equations.

Thus, to construct a physical geometry one needs to express all propositions
of the proper Euclidean geometry in terms of the Euclidean world function $%
\sigma _{\mathrm{E}}$.

\section{Non-Euclidean method of the physical \newline
geometry construction (deformation principle)}

Any physical geometry is described by the world function and obtained as a
result of deformation of the proper Euclidean geometry. The world function
is described by the relation%
\begin{equation}
\sigma :\qquad \Omega \times \Omega \rightarrow \mathbb{R},\qquad \sigma
\left( P,Q\right) =\sigma \left( Q,P\right) ,\qquad \sigma \left( P,P\right)
=0,\qquad \forall P,Q\in \Omega  \label{a1.1a}
\end{equation}

The vector $\mathbf{PQ}\equiv \overrightarrow{PQ}$ is the ordered set of two
points $\left\{ P,Q\right\} ,$ $P,Q\in \Omega $. The length $\left\vert 
\mathbf{PQ}\right\vert _{\mathrm{E}}$ of the vector $\mathbf{PQ}$ is defined
by the relation 
\begin{equation}
\left\vert \mathbf{PQ}\right\vert _{\mathrm{E}}^{2}=2\sigma _{\mathrm{E}%
}\left( P,Q\right)  \label{a1.2}
\end{equation}%
where index "E" means that the length of the vector is taken in the proper
Euclidean space.

The scalar product $\left( \mathbf{P}_{0}\mathbf{P}_{1}.\mathbf{P}_{0}%
\mathbf{P}_{2}\right) _{\mathrm{E}}$ of two vectors $\mathbf{P}_{0}\mathbf{P}%
_{1}$ and $\mathbf{P}_{0}\mathbf{P}_{2}$ having the common origin $P_{0}$ is
defined by the relation 
\begin{equation}
\left( \mathbf{P}_{0}\mathbf{P}_{1}.\mathbf{P}_{0}\mathbf{P}_{2}\right) _{%
\mathrm{E}}=\sigma _{\mathrm{E}}\left( P_{0},P_{1}\right) +\sigma _{\mathrm{E%
}}\left( P_{0},P_{2}\right) -\sigma _{\mathrm{E}}\left( P_{1},P_{2}\right)
\label{a1.3}
\end{equation}%
which is obtained from the Euclidean relation%
\begin{equation}
\left\vert \mathbf{P}_{1}\mathbf{P}_{2}\right\vert ^{2}=\left\vert \mathbf{P}%
_{0}\mathbf{P}_{2}-\mathbf{P}_{0}\mathbf{P}_{1}\right\vert ^{2}=\left\vert 
\mathbf{P}_{0}\mathbf{P}_{2}\right\vert ^{2}+\left\vert \mathbf{P}_{0}%
\mathbf{P}_{1}\right\vert ^{2}-2\left( \mathbf{P}_{0}\mathbf{P}_{1}.\mathbf{P%
}_{0}\mathbf{P}_{2}\right) _{\mathrm{E}}  \label{a1.4}
\end{equation}

The scalar product of two remote vectors $\mathbf{P}_{0}\mathbf{P}_{1}$ and $%
\mathbf{Q}_{0}\mathbf{Q}_{1}$ has the form 
\begin{equation}
\left( \mathbf{P}_{0}\mathbf{P}_{1}.\mathbf{Q}_{0}\mathbf{Q}_{1}\right) _{%
\mathrm{E}}=\sigma _{\mathrm{E}}\left( P_{0},Q_{1}\right) +\sigma _{\mathrm{E%
}}\left( P_{1},Q_{0}\right) -\sigma _{\mathrm{E}}\left( P_{0},Q_{0}\right)
-\sigma _{\mathrm{E}}\left( P_{1},Q_{1}\right)  \label{b1.5}
\end{equation}

The necessary and sufficient condition of linear dependence of $n$ vectors $%
\mathbf{P}_{0}\mathbf{P}_{1}$, $\mathbf{P}_{0}\mathbf{P}_{2}$,...$\mathbf{P}%
_{0}\mathbf{P}_{n}$ is defined by the relation 
\begin{equation}
F_{n}\left( \mathcal{P}^{n}\right) =0,\qquad \mathcal{P}^{n}\equiv \left\{
P_{0},P_{1},...,P_{n}\right\}  \label{a1.7}
\end{equation}%
where $F_{n}\left( \mathcal{P}^{n}\right) $ is the Gram's determinant 
\begin{eqnarray}
F_{n}\left( \mathcal{P}^{n}\right) &\equiv &\det \left\vert \left\vert
\left( \mathbf{P}_{0}\mathbf{P}_{i}.\mathbf{P}_{0}\mathbf{P}_{k}\right) _{%
\mathrm{E}}\right\vert \right\vert =\det \left\vert \left\vert \sigma _{%
\mathrm{E}}\left( P_{0},P_{i}\right) +\sigma _{\mathrm{E}}\left(
P_{0},P_{k}\right) -\sigma _{\mathrm{E}}\left( P_{i},P_{k}\right)
\right\vert \right\vert  \nonumber \\
i,k &=&1,2,...n  \label{b1.8}
\end{eqnarray}

Collinearity $\mathbf{P}_{0}\mathbf{P}_{1}\parallel \mathbf{Q}_{0}\mathbf{Q}%
_{1}$ of two vectors is a special case of the linear dependence. It
described by the relation 
\begin{equation}
\mathbf{P}_{0}\mathbf{P}_{1}\parallel \mathbf{Q}_{0}\mathbf{Q}_{1}:\qquad
\left\vert \left\vert 
\begin{array}{cc}
\left\vert \mathbf{P}_{0}\mathbf{P}_{1}\right\vert _{\mathrm{E}}^{2} & 
\left( \mathbf{P}_{0}\mathbf{P}_{1}.\mathbf{Q}_{0}\mathbf{Q}_{1}\right) _{%
\mathrm{E}} \\ 
\left( \mathbf{Q}_{0}\mathbf{Q}_{1}.\mathbf{P}_{0}\mathbf{P}_{1}\right) _{%
\mathrm{E}} & \left\vert \mathbf{Q}_{0}\mathbf{Q}_{1}\right\vert _{\mathrm{E}%
}^{2}%
\end{array}%
\right\vert \right\vert =0  \label{a1.9}
\end{equation}%
Two vectors $\mathbf{P}_{0}\mathbf{P}_{1}\ $and $\mathbf{Q}_{0}\mathbf{Q}%
_{1} $ are parallel, if%
\begin{equation}
\mathbf{P}_{0}\mathbf{P}_{1}\upuparrows _{\mathrm{E}}\mathbf{Q}_{0}\mathbf{Q}%
_{1}:\qquad \left( \mathbf{P}_{0}\mathbf{P}_{1}.\mathbf{Q}_{0}\mathbf{Q}%
_{1}\right) _{\mathrm{E}}=\left\vert \mathbf{P}_{0}\mathbf{P}_{1}\right\vert
_{\mathrm{E}}\cdot \left\vert \mathbf{Q}_{0}\mathbf{Q}_{1}\right\vert _{%
\mathrm{E}}  \label{a1.10}
\end{equation}%
Two vectors $\mathbf{P}_{0}\mathbf{P}_{1}\ $and $\mathbf{Q}_{0}\mathbf{Q}%
_{1} $ are antiparallel, if%
\begin{equation}
\mathbf{P}_{0}\mathbf{P}_{1}\uparrow \downarrow _{\mathrm{E}}\mathbf{Q}_{0}%
\mathbf{Q}_{1}:\qquad \left( \mathbf{P}_{0}\mathbf{P}_{1}.\mathbf{Q}_{0}%
\mathbf{Q}_{1}\right) _{\mathrm{E}}=-\left\vert \mathbf{P}_{0}\mathbf{P}%
_{1}\right\vert _{\mathrm{E}}\cdot \left\vert \mathbf{Q}_{0}\mathbf{Q}%
_{1}\right\vert _{\mathrm{E}}  \label{a1.11}
\end{equation}

Two vectors $\mathbf{P}_{0}\mathbf{P}_{1}$ and $\mathbf{Q}_{0}\mathbf{Q}_{1}$
are equivalent (equal) $\mathbf{P}_{0}\mathbf{P}_{1}$eqv$\mathbf{Q}_{0}%
\mathbf{Q}_{1}$, if

\begin{equation}
\mathbf{P}_{0}\mathbf{P}_{1}\text{eqv}\mathbf{Q}_{0}\mathbf{Q}_{1}:\qquad
\left( \mathbf{P}_{0}\mathbf{P}_{1}\upuparrows \mathbf{Q}_{0}\mathbf{Q}%
_{1}\right) \wedge \left( \left\vert \mathbf{P}_{0}\mathbf{P}_{1}\right\vert
=\left\vert \mathbf{Q}_{0}\mathbf{Q}_{1}\right\vert \right)  \label{a1.12}
\end{equation}%
or 
\begin{equation}
\mathbf{P}_{0}\mathbf{P}_{1}\text{eqv}\mathbf{Q}_{0}\mathbf{Q}_{1}:\qquad
\left( \left( \mathbf{P}_{0}\mathbf{P}_{1}.\mathbf{Q}_{0}\mathbf{Q}%
_{1}\right) =\left\vert \mathbf{P}_{0}\mathbf{P}_{1}\right\vert ^{2}\right)
\wedge \left( \left\vert \mathbf{P}_{0}\mathbf{P}_{1}\right\vert =\left\vert 
\mathbf{Q}_{0}\mathbf{Q}_{1}\right\vert \right)  \label{a1.12b}
\end{equation}

The property of the equivalence of two vectors in the proper Euclidean
geometry is reversible and transitive. 
\begin{equation}
\text{if\ \ \ }\mathbf{P}_{0}\mathbf{P}_{1}\text{eqv}\mathbf{Q}_{0}\mathbf{Q}%
_{1},\ \ \text{then \ }\mathbf{Q}_{0}\mathbf{Q}_{1}\text{eqv}\mathbf{P}_{0}%
\mathbf{P}_{1}  \label{a2.2}
\end{equation}%
\begin{equation}
\left( \mathbf{P}_{0}\mathbf{P}_{1}\text{eqv}\mathbf{Q}_{0}\mathbf{Q}%
_{1}\right) \wedge \left( \mathbf{Q}_{0}\mathbf{Q}_{1}\text{eqv}\mathbf{R}%
_{0}\mathbf{R}_{1}\right) \text{ }\Longrightarrow \text{\ }\mathbf{P}_{0}%
\mathbf{P}_{1}\text{eqv}\mathbf{R}_{0}\mathbf{R}_{1}  \label{a2.3}
\end{equation}

In general case of physical geometry the equivalence property is
intransitive. The intransitivity of the equivalence property is connected
with its multivariance, when there are many vectors $\mathbf{Q}_{0}\mathbf{Q}%
_{1}$, $\mathbf{Q}_{0}\mathbf{Q}_{1}^{\prime }$, $\mathbf{Q}_{0}\mathbf{Q}%
_{1}^{\prime \prime }$,...which are equivalent to the vector $\mathbf{P}_{0}%
\mathbf{P}_{1}$, but they are not equivalent between themselves.
Multivariance of the equivalence property is conditioned by the fact, that
the system of equations for determination of the point $Q_{1}$ (at fixed
points $P_{0},P_{1},Q_{0}$) has, many solutions, in general. It is possible
also such a situation, when these equations have no solution.

\section{Construction of geometrical objects \newline
in T-geometry}

Geometrical object $\mathcal{O\subset }\Omega $ is a subset of points in the
point set $\Omega $. In the T-geometry the geometric object $\mathcal{O}$ is
described by means of the skeleton-envelope method. It means that any
geometric object $\mathcal{O}$ is considered to be a set of intersections
and joins of elementary geometric objects (EGO).

The finite set $\mathcal{P}^{n}\equiv \left\{ P_{0},P_{1},...,P_{n}\right\}
\subset \Omega $ of parameters of the envelope function $f_{\mathcal{P}^{n}}$
is the skeleton of elementary geometric object (EGO) $\mathcal{E}\subset
\Omega $. The set $\mathcal{E}\subset \Omega $ of points forming EGO is
called the envelope of its skeleton $\mathcal{P}^{n}$. The envelope function 
$f_{\mathcal{P}^{n}}$%
\begin{equation}
f_{\mathcal{P}^{n}}:\qquad \Omega \rightarrow \mathbb{R},  \label{h2.1}
\end{equation}%
determining EGO is a function of the running point $R\in \Omega $ and of
parameters $\mathcal{P}^{n}\subset \Omega $. The envelope function $f_{%
\mathcal{P}^{n}}$ is supposed to be an algebraic function of $s$ arguments $%
w=\left\{ w_{1},w_{2},...w_{s}\right\} $, $s=(n+2)(n+1)/2$. Each of
arguments $w_{k}=\sigma \left( Q_{k},L_{k}\right) $ is the world function $%
\sigma $ of two points $Q_{k},L_{k}\in \left\{ R,\mathcal{P}^{n}\right\} $,
either belonging to skeleton $\mathcal{P}^{n}$, or coinciding with the
running point $R$. Thus, any elementary geometric object $\mathcal{E}$ is
determined by its skeleton $\mathcal{P}^{n}$ and its envelope function $f_{%
\mathcal{P}^{n}}$. Elementary geometric object $\mathcal{E}$ is the set of
zeros of the envelope function 
\begin{equation}
\mathcal{E}=\left\{ R|f_{\mathcal{P}^{n}}\left( R\right) =0\right\}
\label{h2.2}
\end{equation}

Characteristic points of the EGO are the skeleton points $\mathcal{P}%
^{n}\equiv \left\{ P_{0},P_{1},...,P_{n}\right\} $. The simplest example of
EGO is the segment $\mathcal{T}_{\left[ P_{0}P_{1}\right] }$ of the straight
line between the points $P_{0}$ and $P_{1}$, which is defined by the
relation 
\begin{eqnarray}
\mathcal{T}_{\left[ P_{0}P_{1}\right] } &=&\left\{ R|f_{P_{0}P_{1}}\left(
R\right) =0\right\} ,  \label{a5.1} \\
f_{P_{0}P_{1}}\left( R\right) &=&\sqrt{2\sigma \left( P_{0},R\right) }+\sqrt{%
2\sigma \left( R,P_{1}\right) }-\sqrt{2\sigma \left( P_{0},P_{1}\right) }
\label{a5.2}
\end{eqnarray}

Another example is the cylinder $\mathcal{C}(P_{0},P_{1},Q)$ with the points 
$P_{0},P_{1}$ on the cylinder axis and the point $Q$ on its surface. The
cylinder $\mathcal{C}(P_{0},P_{1},Q)$ is determined by the relation 
\begin{eqnarray}
\mathcal{C}(P_{0},P_{1},Q) &=&\left\{ R|f_{P_{0}P_{1}Q}\left( R\right)
=0\right\} ,  \label{g3.1} \\
f_{P_{0}P_{1}Q}\left( R\right) &=&F_{2}\left( P_{0},P_{1},Q\right)
-F_{2}\left( P_{0},P_{1},R\right)  \nonumber
\end{eqnarray}%
\begin{equation}
F_{2}\left( P_{0},P_{1},Q\right) =\left\vert 
\begin{array}{cc}
\left( \mathbf{P}_{0}\mathbf{P}_{1}.\mathbf{P}_{0}\mathbf{P}_{1}\right) & 
\left( \mathbf{P}_{0}\mathbf{P}_{1}.\mathbf{P}_{0}\mathbf{Q}\right) \\ 
\left( \mathbf{P}_{0}\mathbf{Q}.\mathbf{P}_{0}\mathbf{P}_{1}\right) & \left( 
\mathbf{P}_{0}\mathbf{Q}.\mathbf{P}_{0}\mathbf{Q}\right)%
\end{array}%
\right\vert  \label{g3.2}
\end{equation}%
Here $\sqrt{F_{2}\left( P_{0},P_{1},Q\right) }$ is the area of the
parallelogram, constructed on the vectors $\mathbf{P}_{0}\mathbf{P}_{1}$ and 
$\mathbf{P}_{0}\mathbf{Q}$,\textbf{\ }and $\frac{1}{2}\sqrt{F_{2}\left(
P_{0},P_{1},Q\right) }\ $ is the area of triangle with vertices at the
points $P_{0},P_{1},Q$. The equality $F_{2}\left( P_{0},P_{1},Q\right)
=F_{2}\left( P_{0},P_{1},R\right) $ means that the distance between the
point $Q$ and the axis, determined by the vector $\mathbf{P}_{0}\mathbf{P}%
_{1}$, is equal to the distance between $R$ and the axis. Here the points $%
P_{0},P_{1},Q$ form the skeleton of the cylinder, whereas the function $%
f_{P_{0}P_{1}Q}$ is the envelope function.

In the proper Euclidean geometry the cylinder depends only on the axis $%
\mathcal{T}_{\left[ P_{0}P_{1}\right] }$ (or $\mathcal{T}_{P_{0}P_{1}}$),
passing through the points $P_{0}$ and $P_{1}$. It means, that if the point $%
P_{1}^{\prime }\in \mathcal{T}_{\left[ P_{0}P_{1}\right] }$ and $%
P_{1}^{\prime }\neq P_{1}\wedge P_{1}^{\prime }\neq P_{0},$ the cylinders $%
\mathcal{C}(P_{0},P_{1},Q)$ and $\mathcal{C}(P_{0},P_{1}^{\prime },Q)$
coincide in the proper Euclidean geometry. However, the cylinders $\mathcal{C%
}(P_{0},P_{1},Q)$ and $\mathcal{C}(P_{0},P_{1}^{\prime },Q)$ do not
coincide, in general, in an arbitrary T-geometry. It is a result of the
multivariance of the T-geometry, where the straight lines $\mathcal{T}%
_{P_{0}P_{1}}$ and $\mathcal{T}_{P_{0}P_{1}^{\prime }}$ are in general
different, even if $P_{1}^{\prime }\in \mathcal{T}_{P_{0}P_{1}}$.

\textit{Definition.} Two EGOs $\mathcal{E}\left( \mathcal{P}^{n}\right) $
and $\mathcal{E}\left( \mathcal{Q}^{n}\right) $ are equivalent, if their
skeletons are equivalent and their envelope functions $f_{\mathcal{P}^{n}}$
and $g_{\mathcal{Q}^{n}}$ are equivalent. Equivalence of two skeletons $%
\mathcal{P}^{n}\equiv \left\{ P_{0},P_{1},...,P_{n}\right\} \subset \Omega $
and $\mathcal{Q}^{n}\equiv \left\{ Q_{0},Q_{1},...,Q_{n}\right\} \subset
\Omega $ means that 
\begin{equation}
\mathbf{P}_{i}\mathbf{P}_{k}\text{eqv}\mathbf{Q}_{i}\mathbf{Q}_{k},\qquad
i,k=0,1,...n,\quad i<k  \label{a5.4}
\end{equation}%
Equivalence of the envelope functions $f_{\mathcal{P}^{n}}$ and $g_{\mathcal{%
Q}^{n}}$ means that 
\begin{equation}
f_{\mathcal{P}^{n}}\left( R\right) =\Phi \left( g_{\mathcal{P}^{n}}\left(
R\right) \right) ,\qquad \forall R\in \Omega  \label{a5.5}
\end{equation}%
where $\Phi $ is an arbitrary function, having the property%
\begin{equation}
\Phi :\mathbb{R}\rightarrow \mathbb{R},\qquad \Phi \left( 0\right) =0
\label{a5.5a}
\end{equation}

Equivalence of shapes of two EGOs $\mathcal{E}\left( \mathcal{P}^{n}\right) $
and $\mathcal{E}\left( \mathcal{Q}^{n}\right) $ is determined by equivalence
of shapes of their skeletons $\mathcal{P}^{n}$ and $\mathcal{Q}^{n}$, which
is described by the relations

\begin{equation}
\left\vert \mathbf{P}_{i}\mathbf{P}_{k}\right\vert =\left\vert \mathbf{Q}_{i}%
\mathbf{Q}_{k}\right\vert ,\qquad i,k=0,1,...n,\quad i<k  \label{a5.6}
\end{equation}%
and equivalence of their envelope functions $f_{\mathcal{P}^{n}}$ and $g_{%
\mathcal{Q}^{n}}$ (\ref{a5.5}).

Equivalence of orientations of skeletons $\mathcal{P}^{n}$ and $\mathcal{Q}%
^{n}$ in the point space $\Omega $ is described by the relations%
\begin{equation}
\mathbf{P}_{i}\mathbf{P}_{k}\upuparrows \mathbf{Q}_{i}\mathbf{Q}_{k},\qquad
i,k=0,1,...n,\quad i<k  \label{a5.7}
\end{equation}

Equivalence of shapes and orientations of skeletons is equivalence of
skeletons, described by the relations (\ref{a5.4}).

\section{Existence\ of\ geometrical\ objects as physical \newline
objects}

By definition an elementary geometric object $\mathcal{O}_{\mathcal{P}^{n}}$
exists at the point $P_{0}\in \Omega $ in the space-time as a physical
object, if it exists at any time moment at any place of the space-time.
Mathematically it means, that at any point $Q_{0}\in \Omega $ there exists a
geometrical object $\mathcal{O}_{\mathcal{Q}^{n}}$ with the skeleton $%
\mathcal{Q}^{n}$eqv$\mathcal{P}^{n}$. The relation $\mathcal{Q}^{n}$eqv$%
\mathcal{P}^{n}$ means that%
\begin{equation}
\mathbf{P}_{i}\mathbf{P}_{k}\text{eqv}\mathbf{Q}_{i}\mathbf{Q}_{k},\qquad
i,k=0,1,...n,\qquad i<k  \label{a1.19}
\end{equation}%
According to definition of equivalence (\ref{a1.12b}) the equivalence
equation $\mathbf{P}_{i}\mathbf{P}_{k}$eqv$\mathbf{Q}_{i}\mathbf{Q}_{k}$
means two relations 
\begin{equation}
\left( \mathbf{P}_{i}\mathbf{P}_{k}.\mathbf{Q}_{i}\mathbf{Q}_{k}\right)
=\left\vert \mathbf{P}_{i}\mathbf{P}_{k}\right\vert ^{2},\qquad \left\vert 
\mathbf{P}_{i}\mathbf{P}_{k}\right\vert =\left\vert \mathbf{Q}_{i}\mathbf{Q}%
_{k}\right\vert   \label{a1.20}
\end{equation}%
There are $n(n+1)$ equations for determination of $4n$ coordinates of points 
$Q_{1},Q_{2},...Q_{n}$ in the 4-dimensional space-time. The skeleton $%
\mathcal{P}^{n}$ and the point $Q_{0}$ are supposed to be given.

In the case of Minkowski space-time we have only $2n$ relations (instead of $%
n\left( n+1\right) $) for determination of $4n$ coordinates 
\begin{equation}
\left( \mathbf{P}_{0}\mathbf{P}_{k}.\mathbf{Q}_{0}\mathbf{Q}_{k}\right)
=\left\vert \mathbf{P}_{0}\mathbf{P}_{k}\right\vert \cdot \left\vert \mathbf{%
Q}_{0}\mathbf{Q}_{k}\right\vert ,\qquad \left\vert \mathbf{P}_{0}\mathbf{P}%
_{k}\right\vert =\left\vert \mathbf{Q}_{0}\mathbf{Q}_{k}\right\vert ,\qquad
k=1,2,...n  \label{a1.21a}
\end{equation}%
because in the Minkowski space-time

$\left( \mathbf{P}_{0}\mathbf{P}_{i}\text{eqv}\mathbf{Q}_{0}\mathbf{Q}%
_{i}\right) \wedge \left( \mathbf{P}_{0}\mathbf{P}_{k}\text{eqv}\mathbf{Q}%
_{0}\mathbf{Q}_{k}\right) \Longrightarrow \mathbf{P}_{i}\mathbf{P}_{k}$eqv$%
\mathbf{Q}_{i}\mathbf{Q}_{k}$

The structure of equivalence constraints in the Minkowski space-time is
such, that \textit{two} relations 
\begin{equation}
\left( \mathbf{P}_{0}\mathbf{P}_{k}.\mathbf{Q}_{0}\mathbf{Q}_{k}\right)
=\left\vert \mathbf{P}_{0}\mathbf{P}_{k}\right\vert ^{2},\qquad \left\vert 
\mathbf{P}_{0}\mathbf{P}_{k}\right\vert =\left\vert \mathbf{Q}_{0}\mathbf{Q}%
_{k}\right\vert   \label{a1.22a}
\end{equation}%
determine uniquely four coordinates of the point $Q_{k}$, provided the
vector $\mathbf{P}_{0}\mathbf{P}_{k}$ is timelike, i.e. $\left\vert \mathbf{P%
}_{0}\mathbf{P}_{k}\right\vert ^{2}>0$. Thus, in the Minkowski space-time
any geometrical object exists always, as a physical object. If the number of
the skeleton points increases, the number $n\left( n+1\right) $ of
constraints increases faster, than the number $4n$ of coordinates, to be
determined.

In the simplest case, when all $n\left( n+1\right) /2$ vectors $\mathbf{P}%
_{i}\mathbf{P}_{k}$ of the skeleton are timelike, the relation between $%
n\left( n+1\right) $ constraints and $4n$ coordinates is given by the
following table

\[
\begin{array}{cccc}
n & n\left( n+1\right) & 4n & \text{diff} \\ 
2 & 6 & 8 & 2 \\ 
3 & 12 & 12 & 0 \\ 
4 & 20 & 16 & -4%
\end{array}%
\]%
We see that for $n>3$, the number constraints is larger, than the number of
variables to be determined. It appears that existence of complicated
elementary geometrical objects is impossible.

\section{Evolution of geometrical object}

In some cases skeletons of equivalent geometrical objects may form a chain
of identical skeletons. In such cases we shall speak on temporal evolution
of the geometrical object. For instance, let skeletons $\left\{
P_{0}^{\left( l\right) },P_{1}^{\left( l\right) },...P_{n}^{\left( l\right)
}\right\} ,$ $l=...0,1,...$are equivalent in pairs%
\begin{equation}
\mathbf{P}_{i}^{\left( l\right) }\mathbf{P}_{k}^{\left( l\right) }\text{eqv}%
\mathbf{P}_{i}^{\left( l+1\right) }\mathbf{P}_{k}^{\left( l+1\right)
},\qquad i,k=0,1,...n;\qquad l=...1,2,...  \label{a6.1}
\end{equation}%
and besides%
\begin{equation}
P_{1}^{\left( l\right) }=P_{0}^{\left( l+1\right) },\qquad l=...1,2,...
\label{a6.2}
\end{equation}%
If vectors $\mathbf{P}_{0}^{\left( l\right) }\mathbf{P}_{1}^{\left( l\right)
}$ are timelike $\left\vert \mathbf{P}_{0}^{\left( l\right) }\mathbf{P}%
_{1}^{\left( l\right) }\right\vert >0$, one may speak on the temporal
evolution of the geometrical object $\mathcal{O}\left( \mathcal{P}%
^{n}\right) $, which is described by the chain, consisting of equivalent
skeletons $\mathcal{P}^{n}$. In some cases the temporal evolution arises,
even if the vectors $\mathbf{P}_{0}^{\left( l\right) }\mathbf{P}_{1}^{\left(
l\right) }$ are spacelike. However, one may not speak on a temporal
evolution of a geometrical object, if skeletons of the chain are not
equivalent.

\section{Temporal evolution of two-point objects}

We consider some simple examples of temporal evolution of the skeleton,
consisting of two points, in the flat homogeneous isotropic space-time $V_{%
\mathrm{d}}=\left\{ \sigma _{\mathrm{d}},\mathbb{R}^{4}\right\} $, described
by the world function 
\begin{equation}
\sigma _{\mathrm{d}}=\sigma _{\mathrm{M}}+d\cdot \mathrm{sgn}\left( \sigma _{%
\mathrm{M}}\right) ,\qquad d=\lambda _{0}^{2}=\text{const}>0  \label{a6.3}
\end{equation}%
\begin{equation}
\mathrm{sgn}\left( x\right) =\left\{ 
\begin{array}{l}
1,\ \ \text{if}\ \ x>0 \\ 
0,\ \ \ \ \text{if\ \ }x=0 \\ 
-1,\ \ \text{if}\ \ x<0%
\end{array}%
\right. ,  \label{a6.4}
\end{equation}%
where $\sigma _{\mathrm{M}}$ is the world function of the $4$-dimensional
space-time of Minkowski. $\lambda _{0}$ is some elementary length.

The distorted space $V_{\mathrm{d}}$ describes the real space-time better,
than the Minkowski space-time. Description by means of $V_{\mathrm{d}}$ is
better in the sense, that the space-time (\ref{a6.3}) describes quantum
effects, if the distortion constant $d$ is chosen in the form \cite{R91} 
\begin{equation}
d=\frac{\hbar }{2bc}  \label{a6.5}
\end{equation}%
where $\hbar $ is the quantum constant, $c$ is the speed of the light and $b$
is the universal constant, coupling the geometrical length $\mu $ of the
vector $\mathbf{P}_{i}\mathbf{P}_{i+1}$ in the chain of skeletons with the
conventional mass $m$ of the particle, described by this chain%
\begin{equation}
m=b\mu  \label{a6.6}
\end{equation}%
Consideration of distortion taken, in the form (\ref{a6.5}) means a
consideration of the quantum constant as a parameter of the space-time.

The space-time is discrete in the space-time model (\ref{a6.3}). The
space-time is discrete in the sense that there are no timelike vectors $%
\mathbf{P}_{0}\mathbf{P}_{1}$ with $\left\vert \mathbf{P}_{0}\mathbf{P}%
_{1}\right\vert ^{2}\in \left( 0,\lambda _{0}^{2}\right) $ and there are no
spacelike vectors $\mathbf{P}_{0}\mathbf{P}_{1}$ with $\left\vert \mathbf{P}%
_{0}\mathbf{P}_{1}\right\vert ^{2}\in \left( -\lambda _{0}^{2},0\right) $
However, the space-time model (\ref{a6.3}) is not a final space-time
geometry \cite{R91}. The fact is that the relation 
\begin{equation}
\sigma _{\mathrm{d}}=\sigma _{\mathrm{M}}+\frac{\hbar }{2bc}  \label{a6.7}
\end{equation}%
may be not valid for all $\sigma _{\mathrm{M}}>0$. For explanation of
quantum effects, it is sufficient, that the relation (\ref{a6.7}) be
satisfied for $\sigma _{\mathrm{M}}>\sigma _{0}$, where the constant $\sigma
_{0}$ is determined by the geometrical mass of the lightest massive particle
(electron) by means of relation%
\begin{equation}
\sqrt{2\sigma _{\mathrm{d}}}=\sqrt{2\sigma _{\mathrm{0}}+\frac{\hbar }{bc}}%
\leq \mu _{\mathrm{e}}=\frac{m_{\mathrm{e}}}{b}  \label{a6.8}
\end{equation}%
where $m_{\mathrm{e}}$ is the electron mass.

For $\sigma _{\mathrm{M}}<\sigma _{0}$ the form of the distorted world
function may distinguish from (\ref{a6.3}) and have, for instance, the form 
\begin{equation}
\sigma _{\mathrm{di}}=\sigma _{\mathrm{M}}+\frac{2d}{\pi }\text{arctan}%
\left( \sigma _{\mathrm{M}}\right) ,\qquad d=\lambda _{0}^{2}=\text{const}>0
\label{a6.9}
\end{equation}%
The space-time $V_{\mathrm{di}}$ with world function (\ref{a6.9}) takes
intermediate position between the Minkowski space-time and space-time $V_{%
\mathrm{d}}$, described by (\ref{a6.3}). Space-time $V_{\mathrm{di}}$
describes quantum effects as $V_{\mathrm{d}}$, however, the space-time is
not discrete as $V_{\mathrm{d}}$.

If the world function $\sigma \left( x,x^{\prime }\right) $ is given on a
manifold and have derivatives with respect to arguments $x$ and $x^{\prime }$
at the coinciding points $x=x^{\prime }$, the metric tensor $g_{ik}\left(
x\right) $ is defined via derivatives of the world function in the form 
\begin{equation}
g_{ik}\left( x\right) =\left[ -\frac{\partial ^{2}\sigma \left( x,x^{\prime
}\right) }{\partial x^{i}\partial x^{\prime k}}\right] _{x^{\prime
}=x},\qquad i,k=0,1,2,3  \label{a6.10}
\end{equation}
Evaluation of the infinitesimal space-time interval in the space-time (\ref%
{a6.9}) in the inertial coordinate system gives the result%
\begin{equation}
dS^{2}=\left( 1+\frac{2d}{\pi \sigma _{0}}\right) \left( c^{2}dt^{2}-d%
\mathbf{x}^{2}\right)  \label{a6.11}
\end{equation}%
which means space-time interval between two close points $x$ and $x+dx$
appears to be finite at $\sigma _{0}\rightarrow 0$, even if $dx$ is
infinitesimal. The metric tensor, defined by the relation (\ref{a6.11}),
coincides with the metric tensor of the Minkowski space-time to within a
constant factor.

\subsection{Two connected timelike vectors}

Let we have two connected timelike vectors $\mathbf{P}_{0}\mathbf{P}_{1}$
and $\mathbf{P}_{1}\mathbf{P}_{2}$. If $\mathbf{P}_{0}\mathbf{P}_{1}$eqv$%
\mathbf{P}_{1}\mathbf{P}_{2}$ and vector $\mathbf{P}_{0}\mathbf{P}_{1}$ is
given, the vector $\mathbf{P}_{1}\mathbf{P}_{2}$ can be determined. Let
coordinates of points $P_{0},P_{1},P_{2}$ in the inertial coordinate system
be 
\begin{equation}
P_{0}=\left\{ 0,0,0,0\right\} ,\qquad P_{0}=\left\{ s,0,0,0\right\} ,\qquad
P_{2}=\left\{ 2s+\alpha _{0},\gamma _{1},\gamma _{2},\gamma _{3}\right\}
\label{a6.14}
\end{equation}%
where the quantity $s$ is given and the quantities $\alpha _{0},\gamma
_{1},\gamma _{2},\gamma _{3}$ are to be determined.

Vectors $\mathbf{P}_{0}\mathbf{P}_{1}$ and $\mathbf{P}_{1}\mathbf{P}_{2}$
have coordinates%
\begin{equation}
\mathbf{P}_{0}\mathbf{P}_{1}=\left\{ s,0,0,0\right\} ,\qquad \mathbf{P}_{1}%
\mathbf{P}_{2}=\left\{ s+\alpha _{0},\gamma _{1},\gamma _{2},\gamma
_{3}\right\}  \label{a6.15}
\end{equation}%
According to (\ref{b1.5}) in the space-time $V_{\mathrm{d}}$%
\begin{eqnarray}
\left( \mathbf{P}_{0}\mathbf{P}_{1}.\mathbf{P}_{1}\mathbf{P}_{2}\right)
&=&\sigma \left( P_{0},P_{2}\right) -\sigma \left( P_{0},P_{1}\right)
-\sigma \left( P_{1},P_{2}\right)  \nonumber \\
&=&\sigma _{\mathrm{M}}\left( P_{0},P_{2}\right) -\sigma _{\mathrm{M}}\left(
P_{0},P_{1}\right) -\sigma _{\mathrm{M}}\left( P_{1},P_{2}\right) +w 
\nonumber \\
&=&\left( \mathbf{P}_{0}\mathbf{P}_{1}.\mathbf{P}_{1}\mathbf{P}_{2}\right) _{%
\mathrm{M}}+w  \label{a6.16}
\end{eqnarray}%
where%
\begin{equation}
w=\lambda _{0}^{2}\left( \mathrm{sgn}\left( \sigma _{\mathrm{M}}\left(
P_{0},P_{2}\right) \right) -\mathrm{sgn}\left( \sigma _{\mathrm{M}}\left(
P_{0},P_{1}\right) \right) -\mathrm{sgn}\left( \sigma _{\mathrm{M}}\left(
P_{1},P_{2}\right) \right) \right)  \label{a6.17}
\end{equation}%
and $\sigma _{\mathrm{M}}$ means the world function of the Minkowski
space-time.

Note that in the space-time (\ref{a6.3})%
\begin{equation}
\mathrm{sgn}\left( \sigma _{\mathrm{M}}\left( P_{0},P_{2}\right) \right) =%
\mathrm{sgn}\left( \sigma \left( P_{0},P_{2}\right) \right)  \label{a6.18}
\end{equation}%
If the vector $\mathbf{P}_{0}\mathbf{P}_{1}$ is timelike, $\left\vert 
\mathbf{P}_{0}\mathbf{P}_{1}\right\vert ^{2}=s^{2}+2\lambda _{0}^{2}>0,$ The
equivalence equations of vectors $\mathbf{P}_{0}\mathbf{P}_{1}$ and $\mathbf{%
P}_{1}\mathbf{P}_{2}$ take the form%
\begin{equation}
\left\vert \mathbf{P}_{0}\mathbf{P}_{1}\right\vert _{\mathrm{M}%
}^{2}=\left\vert \mathbf{P}_{1}\mathbf{P}_{2}\right\vert _{\mathrm{M}%
}^{2},\qquad \left( \mathbf{P}_{0}\mathbf{P}_{1}.\mathbf{P}_{1}\mathbf{P}%
_{2}\right) _{\mathrm{M}}+w=\left\vert \mathbf{P}_{0}\mathbf{P}%
_{1}\right\vert _{\mathrm{M}}^{2}+2\lambda _{0}^{2}  \label{a6.19}
\end{equation}%
where%
\begin{equation}
w=\lambda _{0}^{2}\left( \mathrm{sgn}\left( \sigma _{\mathrm{M}}\left(
P_{0},P_{2}\right) \right) -2\right)  \label{a6.20}
\end{equation}

Scalar products with a subscript "M" are usual scalar products in the
Minkowski space-time. In the coordinate form the relations (\ref{a6.19}) are
written as follows%
\begin{equation}
\left( s+\alpha _{0}\right) ^{2}-\gamma _{1}^{2}-\gamma _{2}^{2}-\gamma
_{3}^{2}=s^{2}  \label{a6.21}
\end{equation}%
\begin{equation}
s\left( s+\alpha _{0}\right) +\lambda _{0}^{2}\left( \mathrm{sgn}\left(
\sigma _{\mathrm{M}}\left( P_{0},P_{2}\right) \right) -2\right)
=s^{2}+2\lambda _{0}^{2}  \label{a6.22}
\end{equation}%
The vector $\mathbf{P}_{0}\mathbf{P}_{2}=\left\{ 2s+\alpha _{0},\gamma
_{1},\gamma _{2},\gamma _{3}\right\} $ have the length%
\begin{equation}
\left\vert \mathbf{P}_{0}\mathbf{P}_{2}\right\vert _{\mathrm{M}}^{2}=\left(
2s+\alpha _{0}\right) ^{2}-\gamma _{1}^{2}-\gamma _{2}^{2}-\gamma _{3}^{2}
\label{a6.22a}
\end{equation}%
Using relations (\ref{a6.21}) and (\ref{a6.22}), we eliminate $\alpha _{0}$
and $\gamma _{1}^{2}+\gamma _{2}^{2}+\gamma _{3}^{2}$ from the relation (\ref%
{a6.22a}). We obtain%
\begin{equation}
\left\vert \mathbf{P}_{0}\mathbf{P}_{2}\right\vert _{\mathrm{M}%
}^{2}=4s^{2}+8\lambda _{0}^{2}-2\lambda _{0}^{2}\mathrm{sgn}\left( \sigma _{%
\mathrm{M}}\left( P_{0},P_{2}\right) \right) >0  \label{a6.23a}
\end{equation}%
It means that $\mathrm{sgn}\left( \sigma _{\mathrm{M}}\left(
P_{0},P_{2}\right) \right) =1$, and the relation (\ref{a6.22}) has the form 
\begin{equation}
s\alpha _{0}=3  \label{a6.24a}
\end{equation}%
Solution of equations\ (\ref{a6.21}) and (\ref{a6.24a}) has the form%
\begin{equation}
\alpha _{0}=\frac{3\lambda _{0}^{2}}{s},\qquad \gamma _{\alpha }=\frac{%
\lambda _{0}}{s}\sqrt{6s^{2}+9\lambda _{0}^{2}}\frac{\beta _{\alpha }}{\sqrt{%
\beta _{1}^{2}+\beta _{2}^{2}+\beta _{3}^{2}}},\qquad \alpha =1,2,3
\label{a6.24}
\end{equation}%
where $\beta _{1},\beta _{2},\beta _{3}$ are arbitrary real quantities.

Representing coordinates $\left( \mathbf{P}_{1}\mathbf{P}_{2}\right) _{i}$
of the vector $\mathbf{P}_{1}\mathbf{P}_{2}$ in the form%
\begin{equation}
\left( \mathbf{P}_{1}\mathbf{P}_{2}\right) _{i}=\left( \mathbf{P}_{0}\mathbf{%
P}_{1}\right) _{i}+a_{i}  \label{a6.25}
\end{equation}%
where 4-vector $a_{i}$ describes the difference between the vectors $\mathbf{%
P}_{1}\mathbf{P}_{2}$ and $\mathbf{P}_{0}\mathbf{P}_{1}$ in the Minkowski
space-time. We obtain 
\begin{equation}
a_{i}=\left( \frac{3\lambda _{0}^{2}}{s},\frac{\lambda _{0}}{s}\sqrt{%
6s^{2}+9\lambda _{0}^{2}}\frac{\mathbf{q}}{\left\vert \mathbf{q}\right\vert }%
\right)  \label{a6.26}
\end{equation}%
where $\mathbf{q}$ is an arbitrary 3-vector. The 4-vector $a_{i}$ is
spacelike 
\begin{equation}
a_{i}a^{i}=\left( a_{i}a^{i}\right) _{\mathrm{M}}+2\lambda _{0}^{2}\mathrm{%
sgn}\left( \left( a_{i}a^{i}\right) _{\mathrm{M}}\right) =-6\lambda _{0}^{2}
\label{a6.27}
\end{equation}%
If the elementary length $\lambda _{0}\rightarrow 0$, the vector $a_{i}$
tends to zero.

\subsection{Two connected null vectors}

Let us consider now two null connected equivalent vectors $\mathbf{P}_{0}%
\mathbf{P}_{1}$ and $\mathbf{P}_{1}\mathbf{P}_{2}$ ($\left\vert \mathbf{P}%
_{0}\mathbf{P}_{1}\right\vert ^{2}=\left\vert \mathbf{P}_{1}\mathbf{P}%
_{2}\right\vert ^{2}=0$). Using the coordinate representation for vectors $%
\mathbf{P}_{0}\mathbf{P}_{1}$ and $\mathbf{P}_{1}\mathbf{P}_{2}$ 
\begin{equation}
\mathbf{P}_{0}\mathbf{P}_{1}=\left( s,s,0,0\right) ,\qquad \mathbf{P}_{1}%
\mathbf{P}_{2}=\left( s+\alpha _{0},s+\alpha _{1},\gamma _{2},\gamma
_{3}\right)  \label{a6.30}
\end{equation}%
we obtain the following conditions of equivalence%
\begin{equation}
\left( \mathbf{P}_{0}\mathbf{P}_{1}.\mathbf{P}_{1}\mathbf{P}_{2}\right) _{%
\mathrm{M}}+w=0,\qquad \left\vert \mathbf{P}_{1}\mathbf{P}_{2}\right\vert
^{2}=0  \label{a6.31}
\end{equation}%
\begin{equation}
w=\lambda _{0}^{2}\mathrm{sgn}\left( \sigma _{\mathrm{M}}\left(
P_{0},P_{2}\right) \right)  \label{a6.32}
\end{equation}
The relations (\ref{a6.31}) and (\ref{a6.32}) are written in terms of
coordinates%
\begin{equation}
s\left( s+\alpha _{0}\right) -s\left( s+\alpha _{1}\right) =-\lambda _{0}^{2}%
\mathrm{sgn}\left( \sigma _{\mathrm{M}}\left( P_{0},P_{2}\right) \right)
\label{a6.33}
\end{equation}%
\begin{equation}
\left( s+\alpha _{0}\right) ^{2}-\left( s+\alpha _{1}\right) ^{2}-\gamma
_{2}^{2}-\gamma _{3}^{2}=0  \label{a6.34}
\end{equation}%
After simplification we obtain%
\begin{equation}
s\left( \alpha _{0}-\alpha _{1}\right) =-\lambda _{0}^{2}\mathrm{sgn}\left(
\sigma _{\mathrm{M}}\left( P_{0},P_{2}\right) \right)  \label{a6.35}
\end{equation}%
\begin{equation}
\left( \alpha _{0}-\alpha _{1}\right) \left( 2s+\alpha _{0}+\alpha
_{1}\right) -\gamma _{2}^{2}-\gamma _{3}^{2}=0  \label{a6.36}
\end{equation}%
For the length of the vector $\mathbf{P}_{0}\mathbf{P}_{2}$ we obtain 
\begin{eqnarray}
\left\vert \mathbf{P}_{0}\mathbf{P}_{2}\right\vert _{\mathrm{M}}^{2}
&=&\left( 2s+\alpha _{0}\right) ^{2}-\left( 2s+\alpha _{1}\right)
^{2}-\gamma _{2}^{2}-\gamma _{3}^{2}  \nonumber \\
&=&\left( \alpha _{0}-\alpha _{1}\right) \left( 4s+\alpha _{0}+\alpha
_{1}\right) -\gamma _{2}^{2}-\gamma _{3}^{2}  \label{a6.37}
\end{eqnarray}%
By means of relations (\ref{a6.35}), (\ref{a6.36}) the relation (\ref{a6.37}%
) takes the form%
\begin{equation}
\left\vert \mathbf{P}_{0}\mathbf{P}_{2}\right\vert _{\mathrm{M}%
}^{2}=-2\lambda _{0}^{2}\mathrm{sgn}\left( \sigma _{\mathrm{M}}\left(
P_{0},P_{2}\right) \right)  \label{a6.38}
\end{equation}%
The relation (\ref{a6.38}) is fulfilled, only if%
\begin{equation}
\left\vert \mathbf{P}_{0}\mathbf{P}_{2}\right\vert _{\mathrm{M}}^{2}=2\sigma
_{\mathrm{M}}\left( P_{0},P_{2}\right) =0  \label{a6.39}
\end{equation}%
In this case the solution of equations has the form 
\begin{equation}
\alpha _{1}=\alpha _{0},\qquad \gamma _{2}=\gamma _{3}=0  \label{a6.40}
\end{equation}%
Thus, in the case of two connected equivalent null vectors $\mathbf{P}_{0}%
\mathbf{P}_{1}$ and $\mathbf{P}_{1}\mathbf{P}_{2}$ we have 
\begin{equation}
\mathbf{P}_{0}\mathbf{P}_{1}=\left( s,s,0,0\right) ,\qquad \mathbf{P}_{1}%
\mathbf{P}_{2}=\left( s+\alpha _{0},s+\alpha _{0},0,0\right)  \label{a6.41}
\end{equation}%
where $\alpha _{0}$ is an arbitrary real number. The result does not depend
on the elementary length $\lambda _{0}$. It takes place in the Minkowski
space-time also.

\subsection{Two connected spacelike vectors}

Let we have two connected spacelike equivalent vectors $\mathbf{P}_{0}%
\mathbf{P}_{1}$ and $\mathbf{P}_{1}\mathbf{P}_{2}$. If $\mathbf{P}_{0}%
\mathbf{P}_{1}$eqv$\mathbf{P}_{1}\mathbf{P}_{2}$ and vector $\mathbf{P}_{0}%
\mathbf{P}_{1}$ is given, the vector $\mathbf{P}_{1}\mathbf{P}_{2}$ can be
determined. Let coordinates points $P_{0},P_{1},P_{2}$ in the inertial
coordinate system be 
\begin{equation}
P_{0}=\left\{ 0,0,0,0\right\} ,\qquad P_{1}=\left\{ 0,l,0,0\right\} ,\qquad
P_{2}=\left\{ \alpha _{0},2l+\gamma _{1},\gamma _{2},\gamma _{3}\right\}
\label{a6.42}
\end{equation}%
where the quantity $l$ is given and the quantities $\alpha _{0},\gamma
_{1},\gamma _{2},\gamma _{3}$ are to be determined.

Vectors $\mathbf{P}_{0}\mathbf{P}_{1}$ and $\mathbf{P}_{1}\mathbf{P}_{2}$
have coordinates%
\begin{equation}
\mathbf{P}_{0}\mathbf{P}_{1}=\left\{ 0,l,0,0\right\} ,\qquad \mathbf{P}_{1}%
\mathbf{P}_{2}=\left\{ \alpha _{0},l+\gamma _{1},\gamma _{2},\gamma
_{3}\right\}  \label{a6.43}
\end{equation}%
\begin{equation}
\left( \mathbf{P}_{0}\mathbf{P}_{1}.\mathbf{P}_{1}\mathbf{P}_{2}\right)
=\left( \mathbf{P}_{0}\mathbf{P}_{1}.\mathbf{P}_{1}\mathbf{P}_{2}\right) _{%
\mathrm{M}}+w  \label{a6.44}
\end{equation}%
where%
\begin{equation}
w=\lambda _{0}^{2}\left( \mathrm{sgn}\left( \sigma _{\mathrm{M}}\left(
P_{0},P_{2}\right) \right) +2\right)  \label{a6.45}
\end{equation}%
In the coordinate representation the equivalence equations take the form%
\begin{equation}
-l\left( l+\gamma _{1}\right) +w=-l^{2}-2\lambda _{0}^{2},\qquad \alpha
_{0}^{2}-\left( l+\gamma _{1}\right) ^{2}-\gamma _{2}^{2}-\gamma
_{3}^{2}=-l^{2}  \label{a6.46}
\end{equation}%
For the vector $\mathbf{P}_{0}\mathbf{P}_{2}$ we have 
\begin{equation}
\left\vert \mathbf{P}_{0}\mathbf{P}_{2}\right\vert _{\mathrm{M}}^{2}=\alpha
_{0}^{2}-\left( 2l+\gamma _{1}\right) ^{2}-\gamma _{2}^{2}-\gamma _{3}^{2}
\label{a6.47}
\end{equation}%
Eliminating $\gamma _{1}$ and $\alpha _{0}^{2}-\gamma _{2}^{2}-\gamma
_{3}^{2}$ from (\ref{a6.47}) by means of (\ref{a6.46}), we obtain%
\begin{equation}
-l\gamma _{1}+w=-2\lambda _{0}^{2},\qquad \left\vert \mathbf{P}_{0}\mathbf{P}%
_{2}\right\vert _{\mathrm{M}}^{2}=-2l\left( 2l+\gamma _{1}\right)
\label{a6.47a}
\end{equation}%
\begin{equation}
\left\vert \mathbf{P}_{0}\mathbf{P}_{2}\right\vert _{\mathrm{M}%
}^{2}=-2l\left( 2l+\frac{w+2\lambda _{0}^{2}}{l}\right) =-4l^{2}-2\lambda
_{0}^{2}\left( \mathrm{sgn}\left( \sigma _{\mathrm{M}}\left(
P_{0},P_{2}\right) \right) +2\right)  \label{a6.48}
\end{equation}%
It follows from (\ref{a6.48}) that the vector $\mathbf{P}_{0}\mathbf{P}_{2}$
is always spacelike, and, hence, $w=\lambda _{0}^{2}$. It follows from (\ref%
{a6.46}), that 
\begin{equation}
\gamma _{1}=\frac{3\lambda _{0}^{2}}{l},\qquad \alpha _{0}=\sqrt{\gamma
_{2}^{2}+\gamma _{3}^{2}+6\lambda _{0}^{2}+\frac{9\lambda _{0}^{4}}{l^{2}}}
\label{a6.49}
\end{equation}%
where $\gamma _{1}$ and $\gamma _{2}$ are arbitrary real numbers. Thus%
\begin{equation}
\mathbf{P}_{0}\mathbf{P}_{1}=\left\{ 0,l,0,0\right\} ,\qquad \mathbf{P}_{1}%
\mathbf{P}_{2}=\left\{ \sqrt{\gamma _{2}^{2}+\gamma _{3}^{2}+6\lambda
_{0}^{2}+\frac{9\lambda _{0}^{4}}{l^{2}}},l,\gamma _{2},\gamma _{3}\right\}
\label{a6.50}
\end{equation}%
Representing coordinates of the vector $\mathbf{P}_{1}\mathbf{P}_{2}$ in the
form%
\begin{equation}
\left( \mathbf{P}_{1}\mathbf{P}_{2}\right) _{i}=\left( \mathbf{P}_{0}\mathbf{%
P}_{1}\right) _{i}+a_{i}  \label{a6.51}
\end{equation}%
we obtain for the 4-vector $a_{i}$ 
\begin{equation}
a_{i}=\left\{ \sqrt{\gamma _{2}^{2}+\gamma _{3}^{2}+6\lambda _{0}^{2}+\frac{%
9\lambda _{0}^{4}}{l^{2}}},0,\gamma _{2},\gamma _{3}\right\} ,\qquad \left(
a_{i}a^{i}\right) _{\mathrm{M}}=6\lambda _{0}^{2}+\frac{9\lambda _{0}^{4}}{%
l^{2}}  \label{a6.52}
\end{equation}

Let us imagine now that there is an infinite chain of connected equivalent
vectors ...$\mathbf{P}_{0}\mathbf{P}_{1},\mathbf{P}_{1}\mathbf{P}_{2},...%
\mathbf{P}_{k}\mathbf{P}_{k+1},....$If the vectors are timelike, the chain
may be interpreted as a multivariant "world line" of a free particle. The
vector $\mathbf{P}_{k}\mathbf{P}_{k+1}$ may be interpreted as the geometric
particle momentum and $\left\vert \mathbf{P}_{k}\mathbf{P}_{k+1}\right\vert $
may be interpreted as the geometric mass $\mu $. To obtain the conventional
particle mass $m$, one needs to use the relation (\ref{a6.6}), where $b$ is
some universal constant. Statistical description of the multivariant
particle motion leads to quantum description in terms of the Schr\"{o}dinger
equation \cite{R91}. However, this correspondence between the geometrical
description and the quantum one admits one to determine only production $%
\lambda _{0}^{2}b=\hbar /\left( 2c\right) $. The universal constants $%
\lambda _{0}$ and $b$ are not determined asunder from this relation. Thus,
in the case of timelike vector $\mathbf{P}_{k}\mathbf{P}_{k+1}$ we obtain
dynamics of free particle from the pure geometrical consideration (dynamics
is a corollary of geometry).

If the vectors $\mathbf{P}_{k}\mathbf{P}_{k+1}$ of the chain are null, it is
difficult to speak about temporal evolution. Although the chain of null
vectors is single-variant, but vectors of the chain may change their
direction, because the constant $\alpha _{0}$ in (\ref{a6.41}) may have any
sign and module.

At first sight, any temporal evolution in the chain of spacelike vectors $%
\mathbf{P}_{k}\mathbf{P}_{k+1}$ is impossible. It is true, provided there
are no additional constraints on the chain of spacelike vectors. However, if
the skeleton contains more than two points, for instance, $%
P_{0},P_{1},Q_{1},...$ the chain, determined by the points $P_{0},P_{1}$,
may contain additional constraints, generated by additional points $%
Q_{1},... $ of the skeleton. These constraints may be such, that the
spacelike "world line" forms a helix with a timelike axis. If so, the helix
may be interpreted as a world line of a particle, moving with the superlight
velocity along some circle. In this case the mean 4-momentum of the particle
is timelike and directed along the helix axis. The direction of mean
4-momentum does not coincide with that of the instantaneous 4-velocity,
which is spacelike. We see a similar situation in the case of the Dirac
particle, where 4-momentum is a usual timelike vector, whereas the velocity
is always equal to the speed of the light. In the classical approximation
the world line of the Dirac particle has the shape of a helix \cite{R004,R04}%
. The world line of a free particle, having a shape of a helix, may be
explained by the circumstance, that the particle is composite, and in
reality there are two connected particles, rotating around their common
center of inertia. However, in this case one needs to explain the nature of
the interaction, connecting two particles. Such a confinement cannot be
explained by means of dynamics, but it can be explained geometrically as a
temporal evolution, generated by the spatial evolution.

We are not sure, that such a situation may appear in the distorted
space-time (\ref{a6.3}). However, there may exist such a space-time
geometry, where the spatial evolution leads to the temporal evolution. Such
a case is rather unexpected from the viewpoint of the conventional
Riemannian space-time geometry.

\section{Metric force fields}

It is well known, that contorting the Minkowski space-time, we obtain the
curved space-time. The space-time curvature generates the gravitational
field, which is connected with the form of the metric tensor. The curvature
is a special form of the space-time deformation, which does not generate
multivariance of the space-time geometry. The world function $\sigma _{%
\mathrm{R}}$ of a Riemannian space satisfies the equation \cite{S60}

\begin{equation}
\sigma _{\mathrm{R},i}g^{ik}\left( x\right) \sigma _{\mathrm{R},k}=2\sigma _{%
\mathrm{R}},\qquad \sigma _{\mathrm{R},k}\equiv \frac{\partial \sigma _{%
\mathrm{R}}\left( x,x^{\prime }\right) }{\partial x^{k}}  \label{a8.1}
\end{equation}%
The two-point world function of the space-time is determined by the equation
(\ref{a8.1}) and by the metric tensor $g^{ik}\left( x\right) $, given at any
point $x$ of the space-time.

However, if the space-time geometry is multivariant, the world function does
not satisfy the equation (\ref{a8.1}), in general. In this case the metric
tensor $g^{ik}\left( x\right) $ does not determine the world function, in
general. For instance, in the case of world function (\ref{a6.9}) the metric
tensor is given by the relation (\ref{a6.11}). It coincides with the metric
tensor of the Minkowski space-time to within a constant factor. However, in
this case the metric tensor does not determine the world function, because
in this case the world function does not satisfy the equation (\ref{a8.1}).
For $\left\vert \sigma \right\vert \gg \left\vert \sigma _{0}\right\vert $
the world function (\ref{a6.9}) satisfies the equation of the type (\ref%
{a8.1}), which has the form%
\begin{equation}
\sigma _{,i}g_{\mathrm{M}}^{ik}\left( x\right) \sigma _{,k}=\left( 1+\frac{2d%
}{\pi \sigma _{0}\left( 1+\left( \frac{\sigma }{\sigma _{0}+d}\right)
^{2}\right) }\right) ^{2}\frac{2\sigma }{1+\frac{d}{\sigma _{0}}}
\label{a8.2}
\end{equation}%
where $g_{\mathrm{M}}^{ik}\left( x\right) $ is the metric tensor of the
Minkowski space-time.

In the case of the single-variant Riemannian space-time geometry, the
influence of the space-time geometry can be imitated by means of a
gravitational field in the Minkowski space-time. In the case of the
space-time geometry (\ref{a6.9}) one also may say, that the space-time
geometry is imitated by means of some metric fields in the Minkowski
space-time. However, in this case the metric fields form a complex of
fields, which cannot be imitated by a one-point field of the type of metric
tensor. It remains to be unknown, how these fields are described and how
they act on the matter. Formally one may speak about such metric fields and
discuss to what extent such metric fields can imitate interaction between
the elementary particles in microcosm.

Classification of these metric fields and their investigation is very
difficult, if we do not take into account their geometrical origin.
Properties of these fields and their description appear to be very exotic,
because they are generated by the multivariant space-time geometry. For
instance, respectively simple space-time geometry (\ref{a6.9}) is imitated
by principles of quantum mechanics, but not by some force fields, because
the quantum principles take into account the property of multivariance,
whereas the conventional force fields ignore multivariance.

In reality, there exists a very important strategic problem. What is the
starting point for investigation of microcosm phenomena? Now this
investigation is produced on the basis of supposition, that elementary
particles are described by some enigmatic wave functions. They move and
interact in accordance with principles of quantum mechanics. Nobody
understand, what does it mean. Nevertheless the researchers try different
approaches and test them, comparing calculations with experimental data. No
principles, only fitting! A use of space-time geometry is very restricted,
because of imperfection of our knowledge on geometry.

After construction of multivariant geometries it became possible to explain
quantum properties as an appearance of the multivariance of the space-time
geometry. Besides, it becomes possible to set the problem of existence of
geometrical objects in the form of physical bodies. At such conditions it
seems more reasonable at first to investigate capacities of the geometric
approach to the elementary particle theory. At geometric approach we also
have fitting. However, this fitting concerns only a choice of proper
space-time geometry (proper world function). As soon as the world function
is chosen, the fitting ceases. One uses only logical reasonings and
mathematical calculations. High-handedness of the theorist imagination is
restricted.

\end{document}